Commentary

# Rethinking the filter bubble? Developing a research agenda for the protective filter bubble

Jacob Erickson[1]

## Abstract
Filter bubbles and echo chambers have received global attention from scholars, media organizations, and the general public. Filter bubbles have primarily been regarded as intrinsically negative, and many studies have sought to minimize their influence. The detrimental influence of filter bubbles is well-studied. Filter bubbles may, for example, create information silos, amplify misinformation, and promote hatred and extremism. However, comparatively few studies have considered the other side of the filter bubble; its protective benefits, particularly to marginalized communities and those living in countries with low levels of press freedom. Through a review of the literature on digital safe spaces and protective filter bubbles, this commentary suggests that there may be a need to rethink the filter bubble, and it proposes several areas for future research.

## Introduction

Algorithmic recommendation systems (i.e., *recommender systems*) are pervasive throughout the internet and mediate much of an individual's digital experience (Milano et al., 2020). These systems tailor an individual's experience based on implicit or explicit feedback, and they can shape the information that an individual encounters (Milano et al., 2020). While recommender systems have certain conveniences, such as enabling users to find information that they are interested in more easily, they have also led to widespread concerns, such as those concerning privacy (Jeckmans et al., 2013) and fairness (Wang et al., 2023). Digital filter bubbles and echo chambers have also received extensive focus from media researchers, particularly on social media and news websites. There is a widespread concern that algorithmic recommendations create filter bubbles that insulate users from diverse perspectives, sometimes without their knowledge (Geschke et al., 2019; Pariser, 2011).

Despite these concerns, there is reason to believe that not all users view filter bubbles as a negative. Users may, for instance, intentionally want to curate recommendations that promote psychological safety and help them connect to underrepresented users or content (Kanai and McGrane, 2021; Zhao, 2023). Others have challenged the assumptions underlying critical analysis of the filter bubble. For instance, Guess (2021) posited that "requiring of citizens that they continuously engage with challenges to their worldviews fundamentally neglects their autonomy." While filter bubbles have received attention for their adverse influence, comparatively few studies have considered their potential benefits. This commentary reviews the literature on digital safe spaces and protective filter bubbles and proposes a forward-looking research agenda for developing a holistic view of their impact.

## Digital safe spaces

Digital safe spaces are rooted in the concept of the *digital counterpublic*, places where individuals gather to "actively resist hegemonic power, contest majoritarian narratives, engage in critical dialogues, or negotiate oppositional identities" (Hill, 2018). These spaces offer opportunities for individuals in marginalized communities to discuss taboo

[1]Stevens Institute of Technology, Hoboken, NJ, USA

**Corresponding author:**
Jacob Erickson, Stevens Institute of Technology, 1 Castle Point Terrace, Hoboken, NJ 07030, USA.
Email: jericks2@stevens.edu

topics openly, find support, and escape online harassment (Clark-Parsons, 2018; Younas et al., 2020). In repressive political environments, platforms like WhatsApp have provided perceived safety for political expression among users (Ooko, 2023). The underlying commonality among these safe spaces is their intentional construction, offering a potential refuge from harsh online or offline conditions.

## Protective filter bubbles

The features of algorithmic curation make protective filter bubbles distinct from traditional online spaces (Abidin, 2021; Zhao, 2023). While filter bubbles have been viewed critically as insulating users from diverse perspectives, there is a growing awareness that they may serve a protective purpose (Kanai and McGrane, 2021; Randazzo and Ammari, 2023; Zhao, 2023). The existing literature points to several examples of protective filter bubbles, including those supporting feminist groups (Kanai and McGrane, 2021), gay men in China (Zhao, 2023), and political dissidents (Randazzo and Ammari, 2023). While it is an emerging concept, this commentary defines the protective filter bubble as *an algorithmically curated information ecosystem that shields users from threats to psychological and physical safety, including targeted threats such as hate speech, discrimination, and political persecution and generalized threats such as distressing media* (Abidin, 2021; Kanai and McGrane, 2021; Randazzo and Ammari, 2023; Zhao, 2023). In contrast to the traditional digital *safe space*, protective filter bubbles may be emergent phenomena, an "unintentional by-product" of personalized algorithmic systems (Zhao, 2023).

By engaging with marginalized users directly, several studies have found nuanced perspectives, with many users viewing protective filter bubbles as a necessity while also approaching them with caution (Kanai and McGrane, 2021; Randazzo and Ammari, 2023; Zhao, 2023). The protective filter bubble can serve an important function by giving marginalized users a space for discussion and exploration. For instance, Randazzo and Ammari (2023) found that algorithmic recommendations helped give voice to trauma survivors by introducing them to the concepts to recognize their experience and the language necessary to express them. The protective filter bubble may be especially important in information environments that are restricted due to political or social repression (Makhortykh and Wijermars, 2023; Zhao, 2023). For instance, Makhortykh and Wijermars (2023) suggest that in countries with low press freedom, such as Russia, algorithmic personalization and protective filter bubbles may facilitate independent and dissident thought.

### Possible research directions

The prior scholarly work on protective filter bubbles provides a path for developing a forward-looking research agenda. In this section, several overlapping research areas will be covered, starting with those that are presently receiving attention but still necessitate further scholarship. The discussion will then progress toward areas of research that are, at present, less developed.

**Looking Outside the West:** As noted previously, the study of filter bubbles has predominantly focused on countries in a Western context, where researchers may generally assume information freedom and a lack of government restrictions. However, filter bubbles have been understudied in the global context. Algorithmic recommendations in countries with low press freedom may create an environment for political or social dissidence (Makhortykh and Wijermars, 2023). While a repressive political environment makes these spaces more challenging to develop, due to the personalized nature of algorithmic recommendations, they may not be easily detected by surveillance (Makhortykh and Wijermars, 2023). Understanding how these spaces develop, how individuals use them to express unfavored ideas and identities, and the impact of their use is worth deeper consideration.

**Protecting Marginalized Groups:** Representations of marginalized and stigmatized groups can influence the self-perceptions of individuals belonging to these groups (Zhang and Haller, 2013). Accordingly, members of marginalized groups may benefit from an information environment that privileges supportive content and minimizes discrimination or bias. While heterogeneous information exposure is usually considered positive, stigmatized individuals may benefit from a more tailored informational environment. For instance, exposure to hateful content can have detrimental effects on individuals belonging to stigmatized groups (Stefanita and Buf, 2021). Recommender systems and filter bubbles may mitigate these issues under certain circumstances by shielding users from hatred while encouraging the formation of support communities. Gaining a better understanding of the role of protective filter bubbles in fostering these communities should be considered another area worthy of present and future consideration.

**Wellness of the General Public:** The information environment can influence individual mental health and wellness. For example, exposure to negative news can provoke negatively valenced emotions such as anxiety or outrage, which may lower general wellness (Johnston and Davey, 1997; Kellerman et al., 2022). Nevertheless, individuals may inadvertently promote this information due to a bias toward negative news consumption (Trussler and Soroka, 2014). Personalized recommendations may further reinforce this behavior. Future research should consider how intentionally designed algorithms can mitigate these effects and seek to understand user perceptions of such interventions. It is vital to explore the extent to which users are aware of the potential impact their online interactions, such as likes, comments, and views, have on their mental well-being. Additionally, assessing whether the implementation of algorithmic content warnings could support users in making informed decisions about the content they consume is a worthwhile endeavor. It is critically important to understand user attitudes toward these interventions and identify the types of systems they would trust in evaluations.

**Developing Protective Recommender Systems:** Concerns about privacy and fairness are common in recommender system research (Jeckmans et al., 2013; Wang et al., 2023). However, deliberate design efforts aimed at fostering psychologically safe and protective environments in recommendations require further study. There are several promising research directions within this space. For instance, currently, "trigger warnings" are often implemented in an ad-hoc manner, though researchers are considering how to automate this process (Stratta et al., 2020; Wiegmann et al., 2023). Recommender systems could be made to weigh the likelihood of adverse psychological effects of content based on a broad set of features, perhaps by weighing factors such as emotional intensity and topical area. This type of measure could be applied in a variety of contexts; for example, Stray (2022) proposed the idea of incorporating a measure of affective polarization in a political context (e.g., the "feeling thermometer") into recommender systems to consider how likely a post or comment would be to contribute to polarized conversations. Further development is needed in this area.

Furthermore, in the algorithmic fairness literature, researchers have quantified "fairness" metrics (sometimes conceptualized as "unfairness" metrics) (Wang et al., 2023) and weighed them alongside a standard objective function. The same logic may be extensible to the concept of the protective filter bubble. For example, a "protective" metric could be maximized alongside other metrics to account for the risk profile that an algorithm presents to users.

**Changing Curation:** How a protective filter bubble manifests is dynamic and affected by the changing nature of social media content curation. Platforms like Facebook have cultivated feeds in the contemporary social media landscape for well over a decade. It is unclear how much of a modern user's recommendations are "new" and coming from outside their filter bubble and how much is being curated from their cultivated list of friends and pages. Quantifying this "new" content is crucial in understanding the contours of the algorithmic filter bubble. Factors such as these are likely to vary across platforms based on factors including platform age and their general recommendation architectures.

**(Harmful) Visibility:** Attention and visibility are often vital to social media "success," and a lack of it can cause consternation and a feeling of marginalization (Zeng and Kaye, 2022). Nevertheless, visibility can also threaten those involved when the cultural or political environment is unfavorable. Making sensitive content visible to those who would benefit from it while keeping it out of the view of those wishing to weaponize it is a crucial concern of the digital filter bubble and of visibility moderation more broadly (Zeng and Kaye, 2022; Zhao, 2023).

Each of the aforementioned research areas works together to form part of a broader research agenda. A fundamental challenge is understanding how users interact with, understand, and experience algorithms (Hargittai et al., 2020). There is a need for foundational research exploring how users experience and find themselves—either through intentional action or inadvertently—in a protective filter bubble. Furthermore, there is a need for algorithms and recommender systems that weigh the different trade-offs that users may wish to make and that they often do on social platforms. By gaining a more nuanced understanding of user perceptions and intentionally working to build systems that support them, digital environments can better serve the needs of a diverse mixture of users.

## Caution

As Kanai and McGrane (2021) observed, maintaining a protective filter bubble may be a substantial undertaking that is unsustainable in the long run for members of marginalized groups. This dynamic will be especially pronounced if the onus is put on individuals rather than algorithms or those who develop them. Furthermore, because algorithms are often a "black box" to users, a filter bubble's actual protection may be less than what is perceived (Zhao, 2023). Misconceptions about a filter bubble's protective qualities could expose marginalized individuals to severe threats. Similarly, algorithms are also liable to change. What may be protective in the current moment may not remain so, and there is a degree of precarity in relying on these systems. Furthermore, Marwick (2023) noted that recommender systems sometimes undermine their protective qualities, for example, by "leaking information" that was not intended to be shared with others through features such as "people you may know" on Facebook. Finally, while the evidence of their negative influence is mixed, filter bubbles present legitimate risks (Bryant, 2020; Pfetsch, 2018). When and where the risks outweigh the benefits is an open question worthy of further research. Filter bubbles can have adverse impacts, such as contributing to a fracturing public sphere (Pfetsch, 2018) or promoting hatred and extremism (Bryant, 2020). It is not that filter bubbles do not present any risks; the preposition discussed herein is that they may not be intrinsically damaging, and users may perceive them as providing valuable barriers.

## Conclusion

While acknowledging filter bubbles' potential risks and pitfalls, this commentary posits that the potential benefits—either real or perceived—are understudied. Current research suggests that filter bubbles may benefit marginalized communities by providing a refuge where individuals can gather to seek support, avoid hatred, and express themselves freely. Furthermore, the protective filter bubble may benefit those living under repressive conditions who otherwise may have difficulty engaging with certain kinds of material or face surveillance threats. This paper aims to illuminate the path forward for future research by reviewing the literature on digital safe spaces and protective filter bubbles and suggesting areas needing additional scholarly work. Algorithmic recommendations are a significant part of the lived experience of digital natives; it is crucial to understand all sides of their impact.

## Acknowledgments

The author wishes to sincerely thank the editorial team, particularly Dr. Jing Zeng, and the anonymous reviewers for their thoughtful feedback and suggestions.

## Declaration of conflicting interests

The author declared no potential conflicts of interest with respect to the research, authorship, and/or publication of this article.

## Funding

The author received no financial support for the research, authorship, and/or publication of this article.

## ORCID iD

Jacob Erickson  https://orcid.org/0000-0002-9262-9257